ARTICLE    OPEN

# Impact of lattice relaxations on phase transitions in a high-entropy alloy studied by machine-learning potentials

Tatiana Kostiuchenko[1], Fritz Körmann [2,3], Jörg Neugebauer[2] and Alexander Shapeev[1]

Recently, high-entropy alloys (HEAs) have attracted wide attention due to their extraordinary materials properties. A main challenge in identifying new HEAs is the lack of efficient approaches for exploring their huge compositional space. Ab initio calculations have emerged as a powerful approach that complements experiment. However, for multicomponent alloys existing approaches suffer from the chemical complexity involved. In this work we propose a method for studying HEAs computationally. Our approach is based on the application of machine-learning potentials based on ab initio data in combination with Monte Carlo simulations. The high efficiency and performance of the approach are demonstrated on the prototype bcc NbMoTaW HEA. The approach is employed to study phase stability, phase transitions, and chemical short-range order. The importance of including local relaxation effects is revealed: they significantly stabilize single-phase formation of bcc NbMoTaW down to room temperature. Finally, a so-far unknown mechanism that drives chemical order due to atomic relaxation at ambient temperatures is discovered.

## INTRODUCTION

High-entropy alloys (HEAs) are multicomponent alloys consisting of four or more elements in high or even equimolar fractions and crystallize into surprisingly simple lattice structures with randomly dispersed atomic species.[1,2] Many so far discovered HEAs possess extraordinary materials properties such as fcc FeCoNiCrMn with high-cryogenic strength[3] or refractory bcc NbMoTaW HEAs, which reveal auspicious high-temperature mechanical strength.[4] This makes these alloys potential candidates for next-generation technological applications.

A main feature of HEAs is that they form solid solutions. Since alloy properties and materials performance are intrinsically linked to the actual state of chemical ordering, a major part of experimental and theoretical research is devoted to characterize the degree of chemical ordering and to identify order–disorder transitions in these alloys.[1,2] As experimental approaches alone are too time-consuming to cover the vast variety of possible alloy combinations, parameter-free ab initio simulations, typically realized by density functional theory (DFT), have gained rapidly increasing attention as a complementary tool to study various properties of HEAs (see, e.g., ref. [5] and references therein).

For computing chemical ordering and related transitions, even for a single HEA, brute-force DFT simulations are, however, usually limited by the supercell size and number of possible configurations which can be taken into account. Therefore, DFT energetics are typically mapped onto effective interaction models via the cluster expansion (CE) technique[6,7] using, e.g., the structure inverse method (sometimes also denoted as Connolly–Williams approach)[8] or employing perturbational approaches such as the generalized perturbation method (GPM).[9] The latter can be combined with the coherent potential approximation (CPA) and is therefore computationally very efficient. For bcc NbMoTaW, its application revealed an order–disorder transition at 750 K to a B2 ordered state with mixed (Mo,W) and (Nb,Ta) on the two sublattices.[10] Other perturbational method based approaches consistently showed such an incipient B2 ordering when the alloy is cooled down from the solid solution.[11] A limitation of such perturbational approaches is the limited inclusion of local relaxation effects. Indeed, explicit supercell calculations of a B2 ordered state as well as a disordered solid solution revealed that the order–disorder transition temperature can be significantly lower if local lattice relaxations are included in the computations.[12] Recently, Wang et al.[13] extended this approach and considered in total 178 ordered supercell configurations to study the phase stability of bcc NbMoTaW. The B2 ordered structure has been included in the pool of structures and it was found to be stable below 600 K. The limitations of such approaches are, however, that the considered structures must be anticipated, i.e., they must be included in the data pool a priori. Moreover, ideal mixing of elements is assumed not accounting for possible short-range order effects.

The CE can include implicitly local lattice relaxations and can be combined with Monte Carlo (MC) simulations to account for chemical short-range order. CE, however, becomes computationally demanding when studying systems with a large number of chemical elements[14] due to increased combinatorial complexity of possible interatomic interactions. Therefore, in practice, for multicomponent alloys with more than three elements the amount of interactions is typically limited to a few interactions. For example, based on nearest-neighbor pair interactions,[15] also found a B2 ordering at ambient temperatures. Due to these limitations it is, however, not clear whether such an ordering at ambient temperatures is real or an artifact of the underlying approximations (e.g., limited range of interaction parameters). The difficulties of converging a CE for multicomponent alloys has been also recently pointed out by Widom et al.[16] on the example of

[1]Skolkovo Institute of Science and Technology, Skolkovo Innovation Center, Nobel St. 3, Moscow 143026, Russia; [2]Computational Materials Design, Max-Planck-Institut für Eisenforschung GmbH, 40237 Düsseldorf, Germany and [3]Materials Science and Engineering, Delft University of Technology, 2628 CD Delft, The Netherlands
Correspondence: Fritz Körmann (f.h.w.kormann@tudelft.nl) or Alexander Shapeev (a.shapeev@skoltech.ru)

Received: 12 October 2018 Accepted: 10 April 2019
Published online: 01 May 2019

finding the ground state of bcc NbMoTaW[17] where a conventional CE fails.

In the present work, we propose an alternative, highly efficient approach by fitting an accurate active-learning machine-trained potential which is used in subsequent MC simulations. Specifically, we employ a recently proposed "on-lattice" machine-learning interaction model called low-rank potential (LRP),[18] which is capable of including relaxation effects as well as it is capable of accurately representing interactions in multicomponent systems. It is thus well-suited for dealing with a large number of components. As it was shown recently, the LRP requires fewer input structures to reach the same accuracy as a CE approach.[18] The model has only two adjustable parameters: an interaction cutoff radius and the approximation rank controlling the number of free parameters. This allowed us to fit such models semiautomatically by active-learning techniques. The approach can take local relaxation effects implicitly into account in a similar spirit as the CE (by allowing for local relaxations in the DFT calculations for the input structures, see Section "Methods"). It also allows for a systematic estimation of errors in the predictions. We demonstrate the power of this new approach to efficiently and accurately explore huge configuration spaces by studying the finite-temperature phase stability of bcc NbMoTaW. Based on these studies we reveal a hitherto not reported chemical ordering at ambient temperatures. The impact atomic relaxations have on the phase stability and short-range order are discussed.

## RESULTS

An advantage of the approach based on machine-learning potentials, which will be introduced below, is that similarly to the CE formalism the configurations used to fit the potential can be chosen from static or fully relaxed (i.e., including local relaxation effects) calculations. In this way one can straightforwardly "switch on" and "off" the impact of local distortions and study their impact on phase stability and short-range order parameters.

In order to construct a machine-learning potential, we first generate a DFT dataset for fitting the potential as shown in the proposed workflow in Fig. 1. The initial training set consisted of 200 randomly generated configurations for which ground state DFT calculations were performed, each configuration constructed from a $2 \times 2 \times 2$ bcc supercell (with 16 atoms) with randomly distributed atomic species. An initial LRP[18] is then fitted on the initial training set (details given in Section "Methods"). In a nutshell, the LRP assumes a partitioning of the energy into contributions of each atomic environment, e.g., for a bcc lattice each environment is defined by nine atoms, i.e., a central atom and its eight nearest neighbors. This partitioning is different from those used in other formalism such as CE, which assumes a partitioning of the energy into individual two-body, three-body, etc., clusters.

For a given random 4-component alloy there are, in principle, $4^9 \approx 250{,}000$ possible atomic environments, however, a model with 250,000 free parameters is impractical to fit. In order to reduce the number of fitting parameters, the LRP makes a specific low-rank assumption on the representation of the interaction[19] (see Section "Methods"). By applying this assumption we find that only about 500 parameters were eventually required to fit the potential with an accuracy of 1 meV/atom. Although 250,000 environments are prohibitively too many for the purpose of fitting the energy contribution of each environment independently, we can still precompute and store the fitted data for each environment, which requires eventually only 2 megabytes of storage. This leads to an extremely fast energy evaluation which allows us to fit not only one, but an ensemble of ten different potential models. An advantage of having such an ensemble of independent potentials is that not only interaction energies can be predicted, but the ensemble further enables us to estimate the predictive error (i.e., uncertainty) due to approximation of DFT energies with LRP in quantities of interest (such as formation energies) by observing the deviation of the different models. To be precise, we calculate the 95% confidence interval and call it the *model uncertainty* of LRP.

Based on the initial training set we evaluated the accuracy and performed LRP-based MC simulations employing a larger $4 \times 2 \times 2$ supercell including 32 atoms. From these simulations 100 additional, low-energy configurations were chosen, recalculated with DFT and added to the training set. These configurations were about 40 meV lower in energy than the randomly chosen initial training configurations. This procedure has been repeated until the trained potential reaches a predictive error of less than 1 meV/atom. Our retraining (active learning) of the potential also ensures that the configurations span a wide range of compositions, from binaries to quaternaries as well as from chemically ordered to disordered configurations.

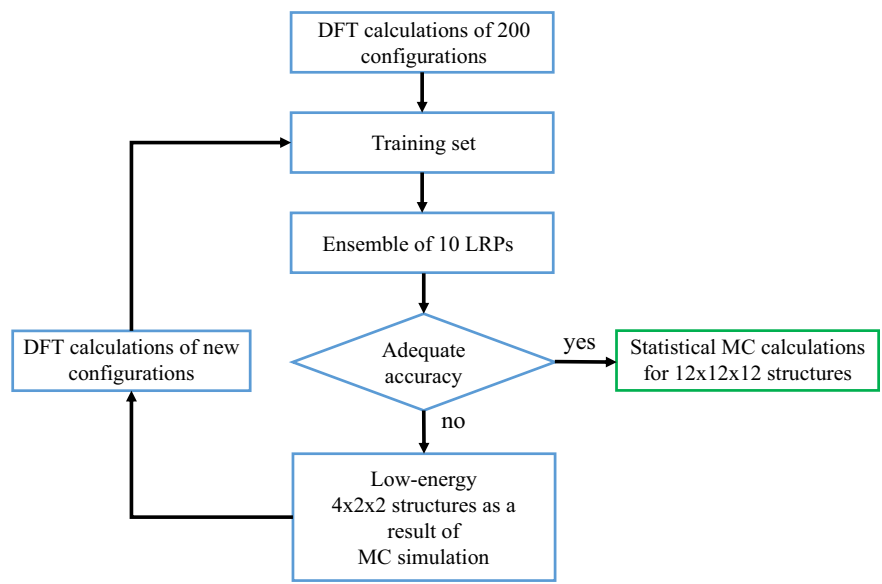

**Fig. 1** The workflow of the proposed approach of HEA investigation

We first exemplify the performance of our proposed approach by training our potentials on a set of static calculations, i.e., with atoms dispersed on the *ideal* bcc positions not including the effect of local lattice relaxations. A set of 10 different potentials are trained and used in MC simulations for a 12 × 12 × 12 supercell to compute the configurational contribution to the heat capacity, $C_V(T)$, and the Warren–Cowley short-range order parameters.[20] We have performed quasiharmonic free energy calculations for a number of selected phases and find that the free energy difference is only about 0.1 meV for $T = 350$ K. Hence vibrational entropy can be safely neglected for the studied phase transitions. The results for $C_V$ are shown in Fig. 2 (blue solid line) and the predictive error is indicated in orange. First we note that the LRP model uncertainty is negligible for a large temperature range. Only close to the two observed phase transitions significant uncertainty due to the different LRPs is present. Note that we have chosen the number of MC steps sufficiently large so that the MC convergence errors are negligible, see Section "Methods". In order to evaluate our approach we first compare the results to previous works which were based on perturbational theories such as the GPM[10] and also employed an ideal lattice approximation. In agreement with ref. [10] we find two phase transitions that the solid solution undergoes with decreasing temperature. The first transition when cooling down corresponds to a transition from the solid solution to a B2(Mo,W;Ta,Nb) ordered state as sketched in Fig. 2. We would like to emphasize that this B2-ordering has also been reported in all previous works so far.[10–13,15,17,21,22] Consistent with ref. [10] a second transition at low temperatures is identified as a phase decomposition into B2(Mo,Ta) and B32(Nb,W) (also sketched in Fig. 2).

The 30% decrease in the phase transition temperatures predicted by LRP as compared to[10] might also affect other properties such as the predicted degree of chemical short-range order at elevated temperatures. To elucidate this we focus next on the high-temperature short-range order parameters

$$a_{ij} = 1 - \frac{p_{ij}}{c_i c_j}, \qquad (1)$$

where $p_{ij}$ is the probability of finding a *j*-type atom among neighbors of *i*-type atoms and $c_i$, $c_j$ are the alloy concentrations of the corresponding elements. The results for the different $a_{ij}$ are shown in Fig. 3. Indeed, the supercell based calculations (dashed lines) reveal overall a smaller degree of short-range order at a given temperature as compared to the ones derived from the perturbation method (open diamonds). We find, however, that the main deviations are caused by the lower transition temperatures predicted by the presently used supercell-based approach, whereas the overall temperature dependencies of the $a_{ij}$'s are rather similar. We note that the SRO for the Mo–Ta bonds is an order of magnitude larger than that for the other bonds as also observed in.[10,15] The question remains whether the assumption of static atomic positions (neglect of lattice relaxations) in these results causes any further qualitative change in the SRO predictions.

In order to account for local lattice distortions, we allowed for ionic relaxations in the configurations entering the training set. As we focus in the present work on the stability of the solid solution (homogeneous disordered alloy) we kept the cell volume and shape fixed to closer mimic the dominant randomized configurations. Note that if our objective was the zero-temperature ground state search, where phase decomposition and thus strong macroscopic volume fluctuations are anticipated, it might be important to consider volume relaxations. As mentioned above, the advantage of the employed LRPs is that it is straightforward to include local lattice distortions into the model. We hence fitted a new ensemble of ten potentials to the DFT data based on relaxed configurations and recomputed the SRO parameters.

The results for the newly trained LRPs, including relaxation effects for the SRO are also shown in Fig. 3 (solid lines). Interestingly, at high temperatures above 600 K—the highest phase transition temperature—the impact of relaxations does not strongly affect the SRO parameters, although relaxations decrease the enthalpy by about 10 meV/atom. This indicates that such a decrease is uniform for most random local environments and does not significantly discriminate one local environment against another one. Based on the exponential-type behavior of the SRO parameters at high temperatures we further corroborated this finding by inspecting the dependence of $\log(1 - a_{ij})$ on $1/T$. From the slope at high temperatures, similar linear dependencies are found for both, relaxed and unrelaxed scenarios. Within a simplified isotropic pair-potential picture, this corresponds to similar *effective* pair interaction energies in the alloy at high temperatures. This implies that perturbation approaches on static lattices are indeed an accurate and computationally efficient tool for studying SRO parameters as done in various previous studies for HEAs.[10,11,23–25] The differences between relaxed and unrelaxed scenarios, however, reveal themselves at lower temperatures. Specifically we find that, compared to the unrelaxed scenario, the Ta–Ta and Nb–Mo bonds become stronger, which prevents segregation of Mo and Ta into a separate phase at low temperatures: Mo and Ta are now also strongly coupled with Nb and Ta, respectively.

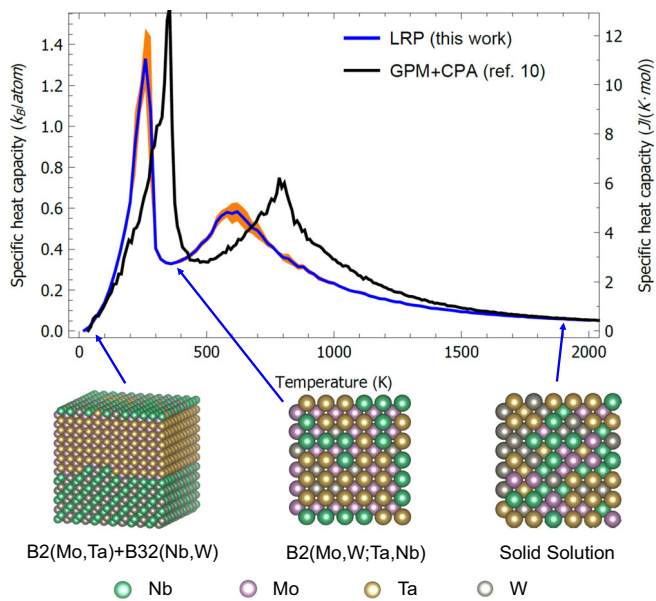

**Fig. 2** The dependence of heat capacity on temperature. The blue line represents the mean heat capacity averaged over the ensemble of LRPs. The standard deviation of the results is shown as the orange area. LRPs predict energy of unrelaxed configurations. The results are presented for the equimolar system with the 12 × 12 × 12 size. The black line represents the data from ref.[10]. The structures correspond to the low-energy structure at 20 K, the ordered layered structure at 360 K and solid solution at high temperature from left to right, respectively

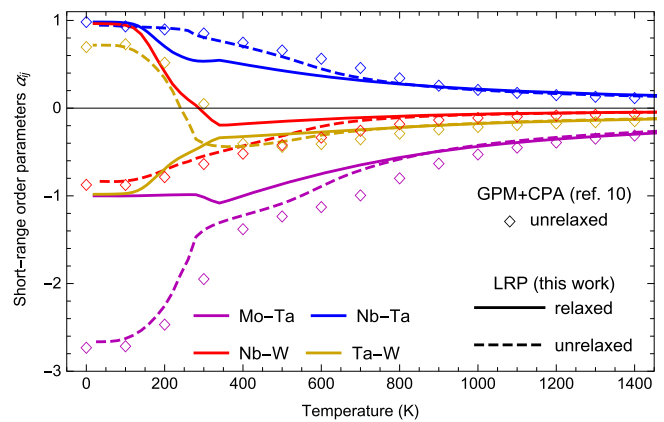

**Fig. 3** The dependence of the short-range order (SRO) parameters on temperature. Dashed lines: LRP using unrelaxed configurations. Solid lines: LRP using relaxed configurations. The results from the perturbative approach[10] are marked by diamonds

To further elaborate the impact of relaxations we computed again the specific heat capacity, now based on the relaxed calculations as shown in Fig. 4. After we introduce the relaxation effects, significant changes can be noticed in the dependence of $C_V(T)$. Still two phase transitions are observed, but now at temperatures of 300 and 200 K, respectively. Relaxation effects have previously been reported to stabilize the solid solution compared to the B2 ordered structure by about 200 K.[12] Here we find that the solid solution is also stabilized by about 200 K, shifting the transition temperature down to 300 K. Note that this is more than 450 K lower than predicted in ref. [10]. The transition temperature is also 200 K lower than predicted in ref. [12], in which the transition temperature has been estimated based on the assumption of ideal mixing and by supercell calculations for the Gibbs energy cross-over of the B2 structure and the solid solution. To examine the origin of these differences we further analyze the structures below the respective transition temperatures.

In our LRP MC simulations below 200 K, an ordered ground-state structure is formed consisting of ⟨100⟩ atomic planes consisting of the same type of atoms. The planes are repeated with the period of four lattice constants in the following sequence: Nb–Mo–Ta–W–W–Ta–Mo–Nb as illustrated in Fig. 4. We stress again that the calculations without relaxations resulted into a phase decomposition into B2(Mo,Ta) + B32(Nb,W) as also reported in ref. [10].

We next analyzed the MC-predicted structures below the first phase transition temperature. We find that the solid solution turns not into a B2 ordered structure (as suggested in all previous works[10–13,15,17,21,22] and our unrelaxed scenario) but into a semiordered layered configuration as sketched in Fig. 4. This new state, which has not been reported before, can be characterized by atomic planes along the ⟨100⟩ direction occupied predominantly by one type of atoms, see Fig. 4. Atoms of other types, if present in such a plane tend to clusterize together into separate islands. A detailed analysis based on large-scale MC simulations including ~200,000 atoms (see also Fig. 5) shows that this phase can be qualitatively described as the repetition of Nb–Mo–Ta–W and W–Ta–Mo–Nb stacks of planes with occasionally appearing W–Ta–Mo–Ta–W and Nb–Mo–Ta–Mo–Nb stacks in such a way that Nb and W planes do not appear near each other. The latter is consistent with the fact that the Nb-W SRO parameters is almost 1 as shown in Fig. 3. This is also consistent with refs. [10,26] where NbW is found to stabilize in a B32 structure suggesting that the second pair interaction energies are larger than the nearest-neighbor ones. This promotes a separation of Nb and W planes. There is a larger number of Mo- and Ta-dominant planes, and the equimolarity is preserved due to higher concentration of Nb and W atoms in the Mo- and Ta-dominant planes than that of Mo and Ta atoms in the Nb- and W-dominant planes. Hence there is a long-range order along these planes in the ⟨100⟩ direction, but no long-range order perpendicular to the planes, as the calculation shown in Fig. 5 confirms.

In order to understand the qualitative difference in the results of calculations with and without taking local lattice distortions into account, we compared the energies of the predicted structures before and after relaxation in DFT calculations. For the structures arising in the unrelaxed scenario, relaxation effects are not significant: the B2(Mo;Ta) and B32(Nb;W) are already at equilibrium due to the symmetry, while the energy of B2(MoW;NbTa) structures decreases by only 0.1 meV/atom after relaxation. On the contrary, the energy of the random phase decreases by approximately 13 meV/atom, while the energies of the ordered and semiordered phases predicted in the relaxed scenario decrease by around 15 meV/atom. This illustrates why accounting

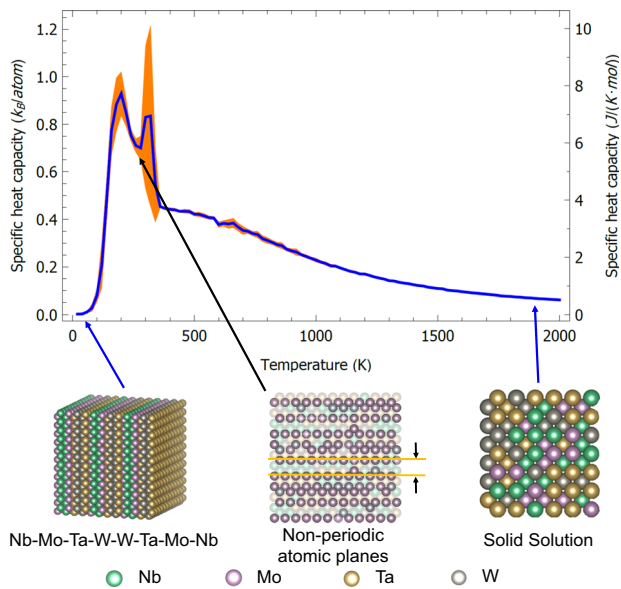

**Fig. 4** The dependence of heat capacity on temperature. The blue line corresponds to the average over an ensemble of LRPs, and the orange area corresponds to the standard deviation. The results are presented for the equimolar system with the 12 × 12 × 12 size. The LRPs predict energy of relaxed configurations. The structures correspond to the low-energy structure at 20 K, the semiordered structure at 240 K with nonperiodically arranged atomic planes and solid solution at high temperature from left to right, respectively. In the semiordered structure, only the Mo atoms are opaque; the other atoms are transparent and form similar planes in other locations

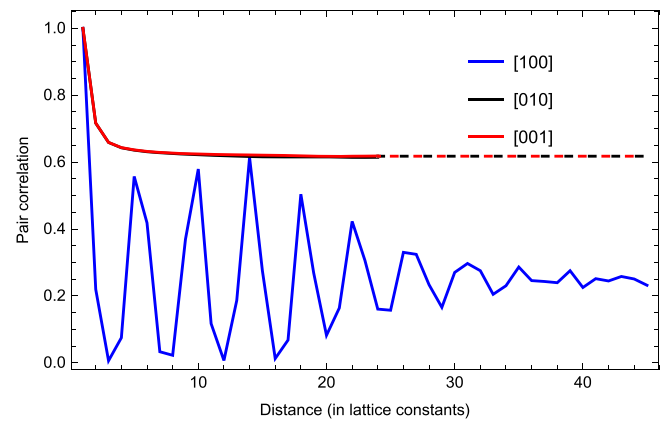

**Fig. 5** Pair correlation for the Mo-Mo pair in the three ⟨100⟩ directions as a function of distance measured in lattice constants. The calculation was done at $T = 240$ K. The planes were formed perpendicular to the [100] direction. The main computation was performed in the 48 × 48 × 48 cell, while the "tail" of the pair correlation function in the [100] direction was computed in the 100 × 20 × 20 cell, as the Monte-Carlo equilibration time grows very fast with the size of the cell. The dashed lines are the extrapolated dependencies

for the lattice relaxation effects is critical for an accurate analysis of the solid solution stabilization of the NbMoTaW alloy.

Finally, we postanalyzed how accurately the LRP model reproduces the energy of the predicted phases. To that end, we have compared with DFT the LRP prediction of the ground state, the semiordered structures, and large $4 \times 4 \times 4$ disordered structures. The average LRP errors for these three phases were 0.1, 1.7, and 0.4 meV/atom, respectively. The low error for the ground state is simply due to the fact that the ground-state structure is added to our training set by our adaptive training procedure. The prediction error for the semi-ordered structure is the highest, but it is still within the 95% confidence interval for LRP. We explain the low prediction error of LRP to the fact that LRP still "saw" the relevant atomic environments in the training set, although the semiordered and large disordered structures were not a part of the training set.

## DISCUSSION

HEAs have emerged as a promising materials class which offers a large compositional phase space for tuning various materials properties. Exploring the vast amount of potential alloys cannot be coped with experiments alone and computational methods are therefore crucial. The underlying key assumption of HEAs is its formation of a solid solution. The solid solution stability range is therefore a critical parameter in processing such alloys and for interpreting their materials properties and performance. The same holds true for the degree of short-range order (SRO) to which many materials properties are intrinsically linked to. To tackle the challenge of accurate phase stability and SRO predictions we proposed a new computational approach. It is based on a combination of density functional theory calculations, machine-learning potentials and subsequent MC simulations. The main advantages of this approach are that it can be combined with supercell calculations including local relaxations as, e.g., the CE, while being extremely efficient and providing a systematic way of improving the accuracy.

Application to a prototypical refractory bcc NbMoTaW alloy revealed the performance of the method. Such refractory alloys are usually prone to limited diffusion and kinetics due to their high-melting temperatures. This renders experimental studies of thermodynamic equilibrium properties at moderate temperatures extremely challenging, further emphasizing the importance of an alternative, computation-driven modeling. Ignoring lattice relaxations, i.e., using a static-lattice approach two phase transitions at 600 and 300 K are observed, namely a B2-ordering at intermediate temperatures and a phase decomposition into B2(Mo,Ta) and B32 (Nb,W) at low temperatures. The two phase transitions including the type of observed ordered configurations agree with the ones observed in previous works where local distortions have been disregarded.[10] The transition temperatures found in the present work are about 150 K lower as compared to the previous reported values. We attribute this in particular to the different employed approximations for mimicking chemical disorder. In the present work supercell calculations have been used whereas in ref. [10] the CPA is employed to simulate chemical disorder and chemical interactions have been extracted by the generalized perturbation method. Moreover only pair interactions have been considered in ref. [10] whereas our approach includes, by construction, also many-body interactions. At high temperatures, the SRO parameters obtained in the present work and ref. [10] are, however, in a good agreement suggesting that perturbation-based approaches such as the GPM or the concentration wave method are powerful tools to study SRO at elevated temperatures.

If local lattice distortions are included in the calculations, our 1 meV/atom-accurate model predicts a sequence of transitions which significantly differ from the results by the static-lattice approach as well as previous simulation results. First, instead of segregation of different species into sublattices we see that the solid solution persists until much lower temperatures as so far suggested, i.e., down to 300 K. Moreover, in contrast to what the previous GPM-based calculations, concentration wave method, and CE-based calculations suggested,[10–13,15,17,21,22] a B2-ordering is not the most stable ordering at ambient temperatures. Indeed, a new, layered semi-ordered phase is found for bcc NbMoTaW. Calculations show that these layered structures gain about −20 meV/atom to their enthalpy after we have allowed for local lattice relaxation and become more energetically favorable by about 5 meV/atom than the phases previously reported as stable —this shows the significance of taking the local lattice relaxation into account. Based on our analysis of machine-learning model uncertainty we suggest that the exact phase transition temperature is in the interval of 250–350 K. As the layered structure is rather different from the originally anticipated B2 ordering, significant impact on various properties such as on the elastic tensor and mechanical anisotropy behavior could be expected. We leave this for future investigations.

At very low temperatures, a layered ordered structure, not reported so far, has been found. Interestingly, Widom has recently revisited bcc NbMoTaW and performed a very careful ground state analysis, and suggested a decomposition into a hR7($Mo_2NbTa_2W_2$) and a cI2(Nb) structure.[16] According to our calculations, its formation enthalpy is indeed by about 2 meV/atom below the low-energy structure we predicted *if* we allow, in addition to the ionic relaxations, for volume changes. This stabilization is mostly driven by pure Nb which has a much larger equilibrium volume than that of the solid solution. If we perform local relaxations only while keeping the volume fixed, the decomposition into hR7($Mo_2NbTa_2W_2$) and cI2(Nb) is about 6.5 meV/atom above the here-found, layered ground state. This also highlights the performance of our approach in predicting the ground state under given external conditions, e.g., at a fixed lattice constant. We would like to emphasize that in contrast to the ground state considerations at zero K, the alloy remains macroscopically homogeneous under the phase transition occurring at room temperature. We, therefore, do not expect that the inclusion of volume fluctuations would qualitatively alter our results at ambient temperatures.

In summary, we have proposed a new computational approach for the investigation of thermodynamic properties of high-entropy alloys. This approach is based on the LRP,[18] a computationally efficient machine-learning interatomic potential capable of accurately representing interactions in a system with many chemical components (see Supplementary Information, Section "LRP predictive power analysis"). The potentials are trained on DFT supercell calculations and thus allow to systematically include the impact of local lattice distortions. The approach is validated by employing a static-lattice approximation and comparing the results to existing ones.[10,12,15] By including local atomic relaxations we found that, contrary to previous works, the solid solution is stable down to room temperature and transforms into a newly found, layered semiordered metastable state. This highlights the important role of local relaxations for the stabilization of the solid-solution where atomic relaxations are not constraint (limited) by symmetry as compared to competing ordered configurations (see Supplementary Information, Section "Analysis of importance of relaxation effects"). An ensemble of potentials further enabled us to analyze the uncertainty of the predictions. The proposed methodology, thus, makes it possible to accurately model multicomponent alloys (including HEAs) in the entire temperature range with high-computational efficiency and to search for new, hitherto unexplored multicomponent ordered states.

# METHODS

## Interatomic potential

In the here-employed approach, the atomic structures are represented by atoms located on fixed lattice sites. Each atom is assigned one of the four atomic types {Nb, Mo, Ta, W}. In this representation local relaxations are allowed, as long as relaxations do not topologically alter the bcc lattice.

In the on-lattice model, LRP,[18] the energy of each configuration is partitioned into contributions of the separate atomic environments of each atom as

$$E(\sigma) = \sum_{\xi \in \Omega} V(\sigma(\xi + r_1), \ldots, \sigma(\xi + r_n)),$$

where $\Omega$ is the lattice periodically repeated in space, $V$ is the so-called interatomic potential defining the contribution of the atom at the lattice site $\xi$ to the total energy $E$, $\sigma(\xi + r_i)$ is the type of $i$th neighbor of the atom located at $\xi$, $r_i$ is the vector connecting $\xi$ with its neighbor, and $n$ is the number of neighbors (including the central atom) that depends on the cutoff parameter.

$V$ can be thought of as a tensor with $m^n$ parameters, where $m$ is the number of atomic types. In order to reduce the number of parameters, the low-rank tensor–train assumption is applied.[19] The tensor–train decomposition of $V$ of rank $r$ is simply

$$V(\sigma_1, \ldots, \sigma_n) = \prod_i A_i(\sigma_i),$$

where $A_i$ are matrices of rank $r$ or less that depend on $\sigma_i \in \{\text{Mo, Nb, Ta, W}\}$. The matrix $A_1$ has the size $1 \times r$, $A_2$ has the size $r \times r$, etc., and $A_n$ has the size $r \times 1$, so that the matrix product results into a scalar. Strictly speaking, it is sufficient to take the size of $A_1$ as $1 \times \min\{4, r\}$, the size of $A_2$ should be $\min\{4, r\} \times \min\{4^2, r\}$, etc. The matrix entries of $A_i$ are the parameters found from data.

The choice of the two adjustable parameters, the number of neighbors $n$ and rank $r$, affect the predictive accuracy of the model. We restricted the interaction to nearest neighbors, $n = 9$, as we found no advantage of considering interaction with longer range. We then kept increasing $r$ until we have reached $r = 5$ that gave us the accuracy of 1 meV/atom. This resulted in about 500 independent parameters.

There can only be $4^n \approx 250{,}000$ possible combination of input parameters $(\sigma_1, \ldots, \sigma_n)$ of $V$. We, therefore, can precompute and store the values of $V$ for all possible inputs. When this is done, the potential $V$ is symmetrized over all 48 permutations of $(\sigma_1, \ldots, \sigma_n)$ corresponding to the physical symmetries. This avoids artificial breaking of symmetry in the MC simulations and slightly improves the accuracy.

In order to find the parameters of $V$ we minimize the following functional:

$$\frac{1}{K} \sum_{k=1}^{K} \left| E(\sigma^{(k)}) - E^{\text{qm}}(\sigma^{(k)}) \right|^2, \qquad (2)$$

where $K$ is the number of the atomic configurations, $\sigma^{(k)}$, in the training set ($k = 1, \ldots, K$), $E(\sigma^{(k)})$ is an energy of configuration predicted by the LRP model, $E^{\text{qm}}(\sigma^{(k)})$ is the DFT reference energy.

The optimization functional Eq. (2) is not linear (because $V$ depends nonlinearly on its tensor–train parameters). Therefore, there are plenty of local minima (by energy) in the space of the parameters. Thus, the minimization algorithm can find different local minima, depending on random initial parameters at the training stage. Therefore, thanks to the fast evaluation of an ensemble of ten different LRPs were trained. Analysis of the independent predictions of the LRP ensemble makes it possible to estimate the uncertainty of the approach.[18]

The workflow of our calculations is shown in Fig. 1. First, we compose the training set from 200 randomly generated configurations, each with 16 atoms. Then the ensemble of 10 LRPs is trained. After this, the accuracy of this ensemble is checked: if the potentials accurately predict the energies of the configurations sampled by the Monte-Carlo algorithm on 32-atom structures, then we can proceed with statistical calculations on the large structures. Otherwise, we add the MC-sampled configurations for different temperatures to the training set, after computing them with DFT. The process hence continues until the accuracy stops improving noticeably as the training set expands. The prediction error (error of the energy prediction of configurations appearing in MC) in our calculations was improved from few meV/atom down to 1 meV/atom by the inclusion of new atomic configurations. We note that the training configurations hence generated contain ordered structures with and without impurity defects, thus ensuring that we take impurities into account together with how they locally distort the lattice.

The configurations (with the reference energy) for the training set are computed with DFT as implemented in VASP 5.4.1. (refs. [27–30]). The lattice constant is set to 3.239 Å which is close to the experimental[4] and theoretical value.[12] A cutoff energy of 400 eV, 1.7 times the default energy cutoff of the PAW pseudopotentials employed, and a dense $8 \times 8 \times 8$ $k$-point mesh generated by the Monkhorst–Pack scheme[31] ($8 \times 8 \times 4$ for the 32-atom configurations) were chosen for the DFT calculations to ensure that the DFT energies are converged down to the error of the order of $10^{-4}$ eV. An additional support grid for the evaluation of the augmentation charges has been employed. Ionic relaxations are performed with a fixed cell volume and shape. The convergence criterion for the electronic loop and the ionic relaxations have been set to $10^{-7}$ and $10^{-6}$ eV, respectively.

## MC method

The canonical MC method is used in the simulations of the equimolar NbMoTaW bcc alloy. The MC simulations are performed with periodic boundary conditions. On-lattice LRPs are used as the interatomic interaction model in the MC scheme.

We studied the temperature range between 20 and 2000 K. Perfect component mixing is guaranteed at 2000 K and nearly perfect ordering is observed at temperatures close to 20 K. Two strategies of temperature changes were considered: (1) starting from 20 K heating the sample by increasing temperature in 20 K steps; (2) starting from 2000 K and cooling down by decreasing the temperature by 20 K at each step. We found that the heating and cooling does not have any significant impact on the evaluated enthalpy, i.e., no hysteresis is observed and both cases coincide. This indicates that our MC algorithm "is ergodic", i.e., explores the full volume of the configurational space typical for a given temperature, because otherwise we would have observed the dependence of the results on the starting configuration through observing hysteresis.

The MC simulations are carried out for a $12 \times 12 \times 12$ supercell based on the 2-atom primitive bcc cell and contains in total 3456 atoms. We have adopted an adaptive number of iterations: For temperatures higher than 600 K, $2 \times 10^7$ MC iterations are performed for each temperature, whereas $2 \times 10^8$ iterations are performed in the temperature range of 400–600 K, and $2 \times 10^9$ iterations are performed for temperatures below 400 K. For the $4 \times 2 \times 2$ cells containing 32 atoms, the number of iterations has been chosen accordingly by two orders of magnitude smaller.

For unbiased averaging, we employed the so-called *burn-in*[32]: we discard half of the MC iterations, and perform averaging over the second half. We note that discarding half of the iterations is necessary only near phase transitions, however, in order to keep our algorithm robust we always discard half of the iterations.

The most challenging MC simulations, for calculating the long-range pair correlation function at $T = 240$ K shown in Fig. 5, were conducted in a $48 \times 48 \times 48$ cell. About $10^{12}$ iterations were required to converge the graphs for the distance of up to 24 lattice constants, however, this was not enough to see the lack of long-range order perpendicular to the planes, in the [100] direction. Calculations in even larger cubic cells, however, are not practical because of a very fast growth of the equilibration time with the cell size. Therefore, in order to see the lack of long-range order in the [100] direction we conducted calculations in a $100 \times 20 \times 20$ cell. Note that these calculations do not provide any new information about the long-range order along the planes, but they reveal the long-range decorrelation perpendicular to the planes.

The atomic structures were visualized with the VESTA software package.[33]

## DATA AVAILABILITY

The datasets generated during and/or analyzed during the current study are available from the corresponding author on reasonable request.

## CODE AVAILABILITY

The code which was used in the findings of this study is available from the corresponding author upon reasonable request.

## ACKNOWLEDGEMENTS

Fruitful discussions with Blazej Grabowski and Marcel H.F. Sluiter are gratefully acknowledged. This collaboration might not have been possible had the authors not met at a number of research programs at the Institute of Pure and Applied Mathematics, UCLA. T.K. and A.S. were supported by the Russian Science Foundation (Grant number 18-13-00479). F.K. acknowledges funding from the Deutsche Forschungsgemeinschaft (SPP 2006) and the Netherlands Organization for Scientific Research NWO/STW (VIDI grant 15707). J.N. acknowledges financial support by the DFG under project number NE 428/19-1.

## AUTHOR CONTRIBUTIONS

F.K. and A.S. designed the research. T.K. and A.S. extended the low-rank potentials to the bcc lattice, F.K. performed the electronic-structure calculations, T.K. and A.S. performed the Monte-Carlo simulations and post-processed the results, all authors discussed the results and wrote the paper.

## ADDITIONAL INFORMATION

**Supplementary Information** accompanies the paper on the *npj Computational Materials* website (https://doi.org/10.1038/s41524-019-0195-y).

**Competing interests:** The authors declare no competing interests.

**Publisher's note:** Springer Nature remains neutral with regard to jurisdictional claims in published maps and institutional affiliations.